\documentclass[twocolumn,a4paper,preprintnumbers,amsmath,amssymb,nofootinbib,floatfix]{revtex4}

\usepackage{color}
\usepackage{graphicx}
\usepackage{dcolumn}
\usepackage{bm}
\usepackage{latexsym}
\usepackage{epsfig}
\usepackage{rotating}
\usepackage{bm}

\newcommand{\be}{\begin{equation}}
\newcommand{\ee}{\end{equation}}
\newcommand{\beqq}{\setlength\arraycolsep{2pt}\begin{eqnarray}}
\newcommand{\eeqq}{\vspace{0cm} \end{eqnarray}}
\newcommand{\bea}{\begin{eqnarray}}
\newcommand{\eea}{\end{eqnarray}}

\begin{document}

\title{Can galaxy clusters, type Ia supernovae and cosmic microwave background rule out a class of  modified gravity theories?}

\author{R. F. L. Holanda$^{1,2,3}$} \email{holanda@uepb.edu.br}
\author{S. H. Pereira$^{4}$} \email{shpereira@gmail.com}

\affiliation{ \\$^1$Departamento de F\'{\i}sica, Universidade Estadual da Para\'{\i}ba, 58429-500, Campina Grande - PB, Brasil,
\\ $^2$Departamento de F\'{\i}sica, Universidade Federal de Campina Grande, 58429-900, Campina Grande - PB, Brasil,\\$^3$Departamento de F\'{\i}sica, Universidade Federal do Rio Grande do Norte, 59300-000, Natal - RN, Brasil.\\$^4$Departamento de F\'isica e Qu\'imica \\ Faculdade de Engenharia de Guaratinguet\'a\\ UNESP - Univ. Estadual Paulista ``J\'ulio de Mesquita Filho''\\  Av. Dr. Ariberto Pereira da Cunha 333 - Pedregulho\\
12516-410 -- Guaratinguet\'a, SP, Brazil.
}



\begin{abstract}
In this paper we study cosmological signatures of modified gravity theories that can be written as a coupling between a extra scalar field and the electromagnetic part of the usual Lagrangian for the matter fields. In these frameworks all the electromagnetic sector of the theory is affected and variations of fundamental constants, of the cosmic distance duality relation and of the evolution law of the cosmic microwave background radiation (CMB) are expected and are related each other. In order to search these variations we perform jointly analyses with angular diameter distances of galaxy clusters, luminosity distances of type Ia supernovae and $T_{CMB}(z)$ measurements. We obtain tight constraints with no {significant} indication of violation of the standard framework. 

\end{abstract}

\maketitle


\section{Introduction}

Since its publication in 1915,  the main theory of gravitation, the General Relativity (GR), has been put in check, since that the observations of galactic velocities in galaxy clusters, the rotational curve of spiral galaxies and the recent discovery of the accelerated expansion of the universe \cite{SN,union2,WMAP,planck,farooq,sharov} via observations of Supernovae type Ia in 1998 only can be  explained correctly with addition of  new ingredients in the nature: the so-called dark matter (DM) and dark energy (DE). The DM is a kind of matter that does not interact electromagnetically with other particles of the standard model \cite{reviewDM,bookDM}. Actually, the DM also has a fundamental role in evolution of cosmic structures in GR context. The DE is an alternative to explain the accelerated evolution of the universe \cite{reviewDE}, once that the cosmological constant (CC), which appears naturally in GR, is plagued with several conceptual problems in which concerns its nature and origin \cite{CC}. In this way, several models of gravity have been appeared in literature in order to give alternatives to GR. Among such alternative models, massive gravity theories \cite{volkov,koba}, modified Newtonian dynamic (MOND) \cite{mond}, $f(R)$ and $f(T)$ theories \cite{fR}, brane world models \cite{randall,pomarol,langlois,shir,orito,cline,okada} among others, have been proposed recently in order to accommodate the observations. On the other hand, it is also important to have mechanisms to test whether these theories actually satisfy various observational constraints.

Recently, a wide class of theories of gravity that explicitly breaks the Einstein equivalence principle (EEP) have been considered in the literature and a powerful mechanism to test its signatures in observable constants of nature has been developed by A. Hees et al. \cite{hees,hees2}. They consider models which implements the break of the equivalence principle by introducing an additional term into the action, coupling the usual matter fields $\Psi$ to a new scalar field $\phi$, which is motivated by scalar-tensor theories of gravity, for instance. The explicit form of the couplings studied by \cite{hees,hees2} are of the type
\begin{equation}
S_{m}=\sum_i \int d^4x\sqrt{-g}h_i(\phi)\mathcal{L}_i(g_{\mu\nu},\Psi_i)\,,\label{action}
\end{equation}
where $\mathcal{L}_i$ are the Lagrangians for different matter fields $\Psi_i$ and $h_i(\phi)$ represents a non-minimal couplings between $\phi$ and $\Psi_i$. When $h_i(\phi)=1$ we recover the standard GR.  Several alternative models can be described by such a kind of coupling. We can cite string dilaton theories \cite{string} at low energies, theories with additional compactified dimensions as Kaluza-Klein \cite{klein}, models involving axions \cite{axion},
cosmologies that consider a varying fine structure constant \cite{fine}, chameleon-field models \cite{chameleon} or $f(R)$ extended gravity theories \cite{fRL}. 

The most direct consequence of a interaction of the type (\ref{action}) concerns its relation to the fine structure constant $\alpha$ of the quantum electrodynamics. It is related to the scalar field $\phi$ by $\alpha \propto h^{-1}(\phi(t))$, such that a time dependence of $\phi$ will leads to a time variation of the fine structure constant  $\alpha$ \cite{fine,const_alpha}. Actually, all the electromagnetic sector of the theory also is affected, which implies in a non-conservation of the photon number along geodesics, leading to a modification to the expression of the luminosity distance, $D_L(z)$, where $z$ is the redshift, which is the basis for various cosmological estimates  and also the violation of the so-called cosmic distance-duality relation (CDDR), $D_L (1+z)^{-2}/D_A=1$, where $D_A$ is the angular diameter distance \cite{distance}. Moreover, also due to the non-conservation of the photon number, it is expected a variation of the  evolution  of the Cosmic Microwave Background (CMB) radiation, affecting its temperature distribution. Finally, as a consequence of the CMB distribution, we also expect a CMB spectral distortion, which can be parametrized by a non-null chemical potential. In \cite{hees} the authors showed that all these effects are closely related to the time evolution of $h(\phi(t))$. By using Gaussian Processes, they also considered  $D_A(z)$ measurements of galaxy clusters obtained via their Sunyaev-Zeldovich + X-ray observations (SZE/X-ray technique), $D_L(z)$ of type Ia supernovae, CMB temperature and absorbers to impose limits on $h(\phi)$. Although the results were not so restrictive,  no inconsistency with the standard results was detected. 

However, in Ref.\cite{colaco} it was showed that the SZE/X-ray technique depends strongly on the CDDR validity as well as on the $\alpha$. These dependencies were used property in Ref.\cite{holandaprd} to search signatures of the equivalence principle breaking. These authors considered the results from Ref.\cite{hees} jointly with galaxy clusters and SNe Ia observations and showed that if the CDDR is not valid, $D_L D_A^{-1}(1+z)^{-2}=\eta$ and $\Delta \alpha/\alpha \neq 1$, the SZE/X-ray technique does not give the true angular diameter distance of galaxy clusters but $D_A^{obs}=D_A\eta^{-3}$. Again, no inconsistency with the standard results was detected by considering two parametrization to $\eta(z)$, such as $\eta(z)=1+\eta_0 z$ and $\eta(z)= 1 + \eta_0 z/(1+z)$. The more restrictive value to $\eta_0$ was $\eta_0=0.069 \pm 0.106$ at 1$\sigma$ c.l..

In this paper we search for signatures of the class of  modified gravity theories discussed by Ref.\cite{hees}  by testing  jointly  the CDDR  and  the evolution law of CMB temperature. We consider angular diameter distance samples from galaxy clusters obtained via the SZE/X-ray technique, luminosity distances from SNe Ia and $T_{CMB}(z)$ measurements. Moreover, four parametrizations for $\eta(z)$ are used. The error bars from our jointly analyses are at least $ 50\%$ smaller than those in Ref.\cite{holandaprd}. {Our results showed no significant deviation from the standard framework ($\eta_0=0$)  regardless the $\eta(z)$ function and galaxy cluster sample used.}

This paper is organized as follows: Section II we briefly revise  the theory proposed by Ref.\cite{hees}. In Section III we present our method and the samples used in analyses. The Section IV we show the analyses and results and finally, in the Section V are the main conclusions of this work.
\section{Consequences of the breaking of the equivalence principle}

As mentioned before, A. Hees et al. \cite{hees,hees2} developed a powerful apparatus to test signatures of models characterized by an interaction term into the action of the form (\ref{action}) in observable constants of nature. Such kind of models implements the break of EEP. We cite here briefly three consequences, namely the temporal variation of the fine structure constant, modification of the CDDR and variations and distortions on CMB temperature.
\subsection{Temporal variation of the fine structure constant}
Having $\alpha \propto h^{-1}(\phi(t))$ \cite{string,fine,DAMOURPIAZZAVENEZIANO,DAMOURPIAZZAVENEZIANO89}, the time variation of the fine structure constant $\alpha$ is associated to time variation of the coupling of the electromagnetic Lagrangian $h_{EM}(\phi)$ by
\begin{equation}
\frac{\dot{\alpha}}{\alpha} =-\frac{h_{EM}^{'}(\phi)}{h_{EM}(\phi)}\dot{\phi}
\end{equation}
where the dot corresponds to the temporal derivative and the prime  to the derivative with respect to the scalar field $\phi$. Writing in terms of the redshift $z$,
\begin{equation}
\frac{\Delta \alpha}{\alpha}=\frac{\alpha (z)-\alpha_0}{\alpha_0}= \frac{h( \phi_0 )}{h( \phi (z))} -1= \eta^2 (z) -1,\label{alpha}
\end{equation}
where the subscript $0$ stands for the present epoch $( \phi_0 = \phi (z = 0))$, we can define the parameter 
\begin{equation}
\eta (z)= \sqrt{\frac{h( \phi_0 )}{h( \phi (z))}}.
\end{equation}\label{etaz}
which can be directly interpreted as a constraint on the cosmological evolution of the scalar field $\phi (z)$. 
\subsection{Modification of the cosmic distance-duality relation}
The expression of the luminosity distance $D_L$ is also modified with respect to the general relativity one \cite{MinazzoliHees}, given by
\begin{equation}
D_L(z)=c(1+z)\sqrt{\frac{h(\phi_0)}{h( \phi (z))}} \int_{0}^{z} \frac{dz'}{H(z')},\label{DL}
\end{equation}
where $c$ is the speed of the light and $H(z)$ is the Hubble parameter. On the other hand, the angular diameter distance $D_A$ is a purely geometric property that is the same as in general relativity and it is given by \cite{HobsonEfstathiouLasenby}
\begin{equation}
D_A(z)=\frac{c}{(1+z)} \int_{0}^{z} \frac{dz'}{H(z')}\,.\label{DA}
\end{equation}
By comparing with (\ref{DL}) we obtain
\begin{equation}
\frac{D_L(z)}{D_A(z)(1+z)^2}= \sqrt{\frac{h( \phi_0 )}{h( \phi (z))}}=\eta (z)\,,\label{DLDA}
\end{equation}
which shows that the CDDR can also be related to $\eta(z)$.

\subsection{Modifications of CMB temperature}

The equations that governs the evolution of the temperature of the CMB are based on the kinetic theory (see it in \cite{PeterUzan} and \cite{Durrer2008}), which satisfy the Boltzmann equations of statistical mechanics. However, a non-conservation of the photon number due to a coupling of the form (\ref{action}) may also alter the evolution of the CMB radiation. There is also a connection between violations of the temperature-redshift relation and variations of the fine structure constant. Furthermore, the coupling (\ref{action}) also implies that the CMB radiation does not obey the adiabaticity condition \cite{LimaSilvaViegas}, whose distortion of the CMB spectrum can be parametrized by a chemical potential $\mu$. The relations for the CMB temperature evolution and the chemical potential as a function of $\eta(z)$ are \cite{hees}
\begin{equation}
T(z)=T_0(1+z)[0.88+0.12 \eta^2(z)], \label{T}
\end{equation}
\begin{equation}
\mu =0.47 \Bigg( 1-\frac{1}{\eta^2 (z_{CMB})} \Bigg)=3.92 \Bigg( \frac{T(z_{CMB})}{T_0(1+z_{CMB})} -1 \Bigg),\label{mu}
\end{equation}
which are related each other. It is useful to express the experimental constraints on the evolution of the temperature as a function of the parameter $\beta$, denoted by
\begin{equation}
T(z)=T_0(1+z)^{1- \beta}.
\end{equation}
A. Hees et al. \cite{hees} have shown that the four cosmological observables, (\ref{alpha}), (\ref{DLDA}), (\ref{T}) and (\ref{mu}) are directly related to the evolution of the function $h( \phi )$ by
\begin{equation}
\frac{h(\phi_0 )}{h(\phi (z))} =\eta^2 (z)=\frac{\Delta \alpha (z)}{\alpha} +1 = 8.33 \frac{T(z)}{T_0 (1+z)} -7.33\,.\label{rel}
\end{equation}
\begin{figure*}[ht]
\centering
\includegraphics[width=0.47\textwidth]{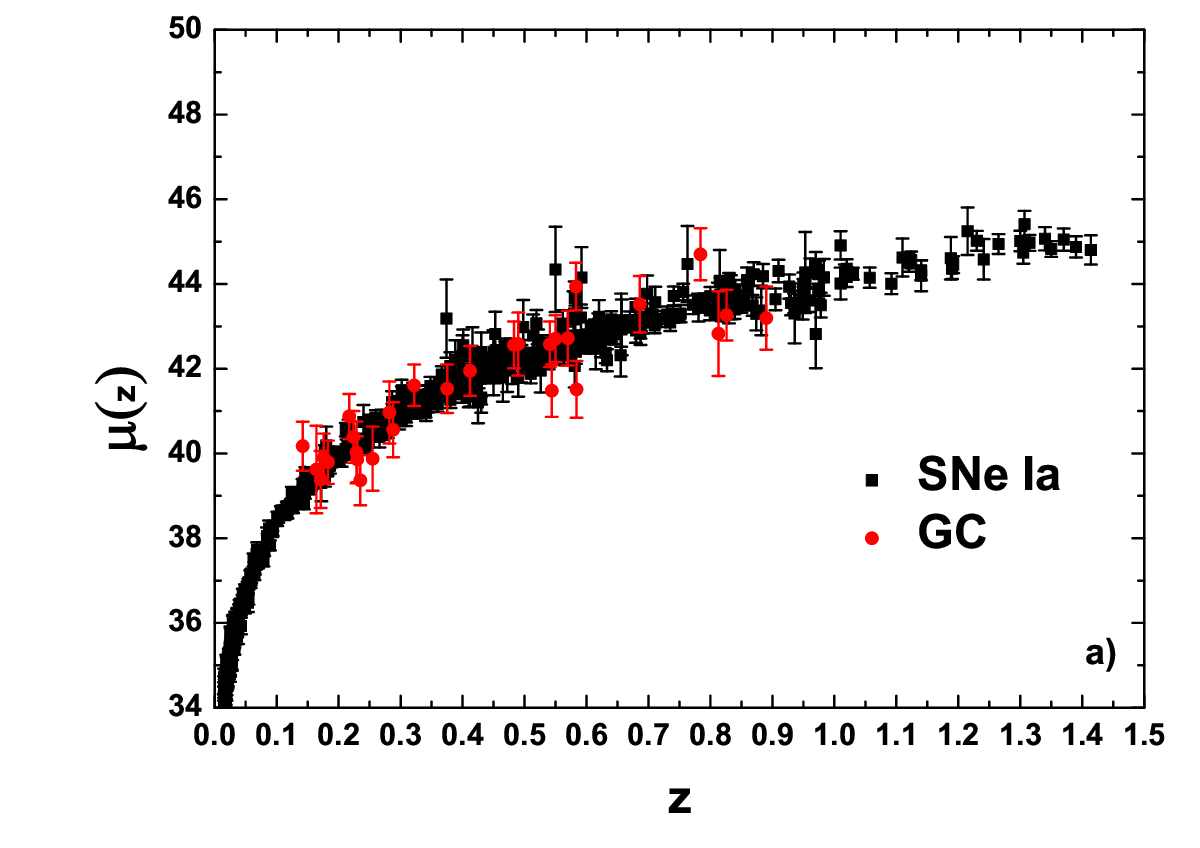}
\hspace{0.3cm}
\includegraphics[width=0.47\textwidth]{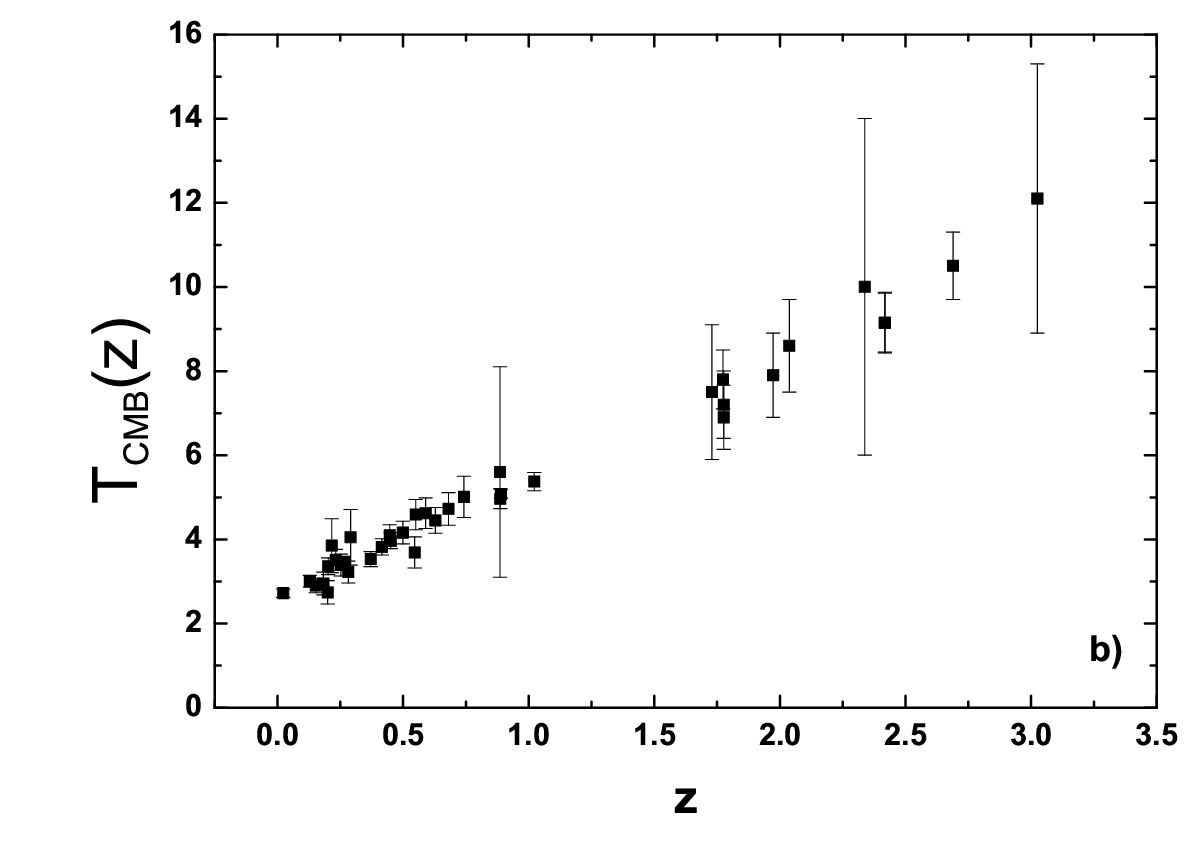}
\caption{ In Fig.(a) We plot the distance modulus of SNe Ia and also for galaxy clusters (GC) by using the Eq.(16) com $\eta=1$. In Fig.(b) we plot $T_{CMB}(z)$ data. }
\end{figure*}
As it is well-known, there are astronomical methods  based  on the analysis of high-redshift quasar absorption systems to test the $\Delta \alpha/\alpha$ value. The many-multiplet method,  which compares the characteristics of different transitions in the same absorption cloud, is the most successful method employed so far to measure possible variations of $\alpha$. However, very recently, constraints on the variations of the fine structure constant $\alpha$ have been derived directly from cosmological observations such as the Sunyaev-Zel'dovich Effect (SZE) and X-ray emission in galaxy cluster. For example, Ref.\cite{Galli2013} proposed a new method using the integrated Comptonization parameter $Y_{SZ} D_{A}^{2}$ and its X-ray counterpart $Y_X$, and the ratio of these two parameters depends on the fine structure constant as $\alpha^{3.5}$. Recently, Holanda et al. \cite{holandaprd} showed that measurements of the gas mass fraction can also be used to probe a possible time evolution of the fine structure constant. For that purpose, they have showed that observations of the gas mass fraction via X-ray surface brightness and the SZE for the same galaxy cluster are related by
\begin{equation}
f_{SZE} = \phi (z) \eta (z) f_{Xray},
\end{equation}
where $\phi (z) = \frac{\alpha}{\alpha_0} $. Taking into account a direct relation between variation of $\alpha$ and the CDDR, Eq. (\ref{DLDA}), and particularizing the analysis by considering a class of dilaton runaway models in which $\phi (z) = 1- \gamma \ln{(1 + z)}$, it was found $\gamma = 0.037 \pm 0.18$ at $1 \sigma$ c.l., consistent with no variation of $\alpha$. More recently, a deeper  analysis from Ref.\cite{colaco} of the SZE/X-ray technique showed that measurements of $D_A (z)$ of galaxy clusters by using this technique also depends on the fine structure constant. They have showed that if $\alpha = \alpha_0 \phi (z)$, current SZE and X-ray observations do not provide the real angular diameter distance but instead
\begin{equation}
D_{A}^{data} (z)= \phi (z) \eta^2 (z)D_A (z).\label{eqnew}
\end{equation}
In order to perform their analysis, they have transformed 25 measurements of $D_L$ from current SNe Ia observations into $D_A (z)$, taking into account the direct relation, shown by Hees et al. \cite{hees}, between a variation of $\alpha$ and the CDDR. When combined with 25 measurements of $D_{A}^{data} (z)$ from galaxy clusters in the range of redshift $0.023 < z < 0.784$, these data sets impose cosmological limits on $\phi (z)$ for a class of dilaton runaway models. So, they have found $\frac{\Delta \alpha}{\alpha} = -(0.042 \pm 0.10)\ln{(1 + z)}$, which is also consistent with no variation of $\alpha$.  On the other hand, in Ref.\cite{holandaprd} it has been used the Eq. (\ref{eqnew}) and the relation between $\alpha$ and $\eta$ to impose tighter limits on deviation from CDDR than previous works. By using  $\eta(z)=1+\eta_0 z$ and $\eta(z)= 1 + \eta_0 z/(1+z)$, the most restrictive value to $\eta_0$ is $\eta_0=0.069 \pm 0.106$ at 1$\sigma$ c.l.. 
\begin{figure*}[ht]
\centering
\includegraphics[width=0.47\textwidth]{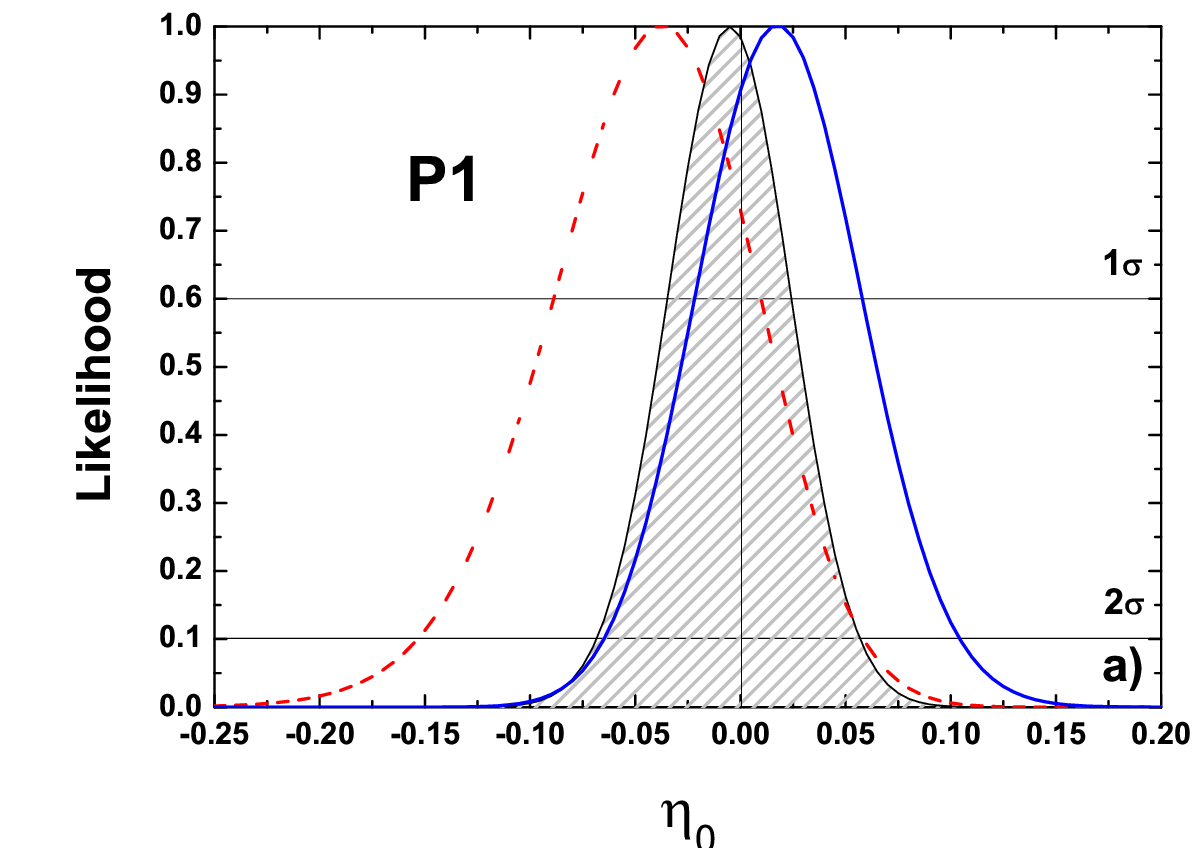}
\hspace{0.3cm}
\includegraphics[width=0.47\textwidth]{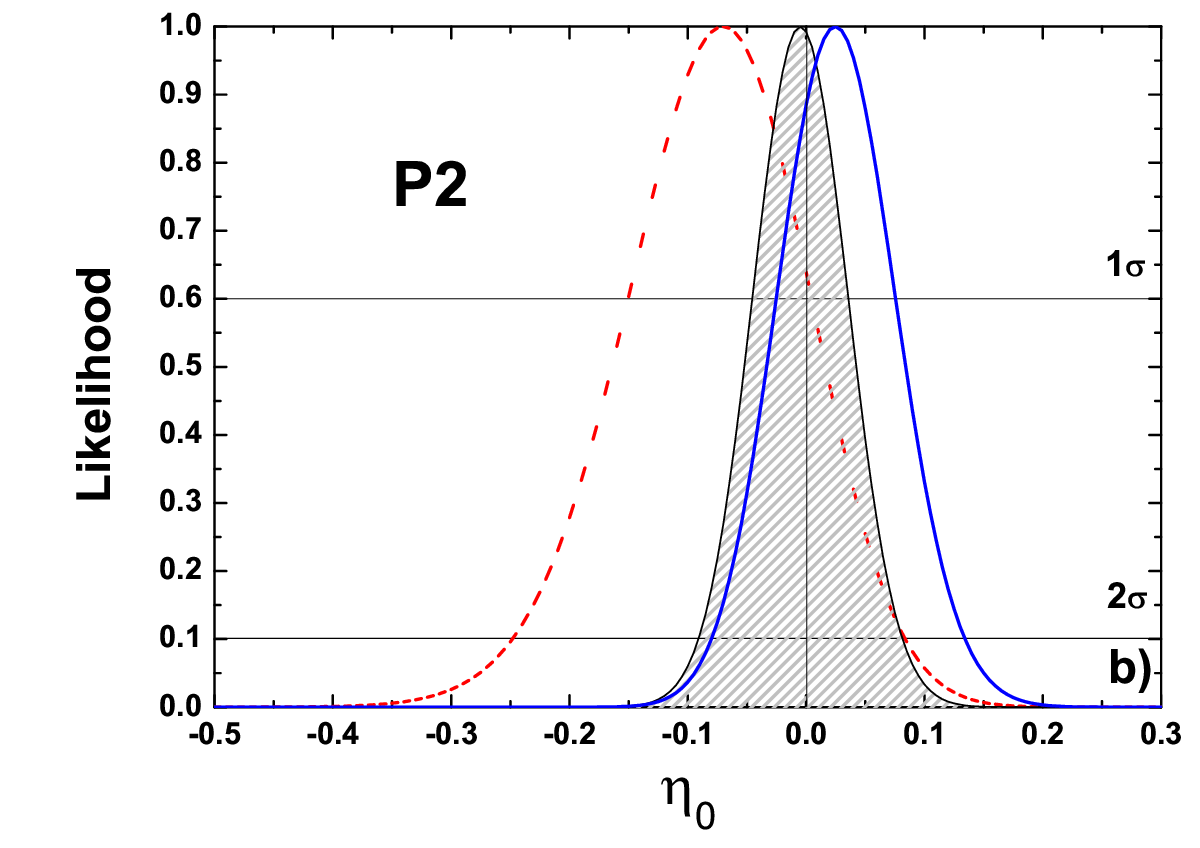}
\caption{In both figures, the solid blue and dashed red lines correspond to analyses by using SNe Ia + GC and  $T_{CMB}(z)$, respectively. The dashed area corresponds to the joint analysis (SNe Ia + GC + $T_{CMB}(z)$). In Fig.(a) we plot the results by using the parametrization P1 and in Fig.(b) by using P2. }
\end{figure*}
\section{New Observational constraints}
In the present paper we test the same class of  modified gravity theories presented in Ref.\cite{hees} following the Ref.\cite{holandaprd}. Nevertheless, we use different  galaxy cluster sample, SNe Ia and we also include measurements of CMB temperature  in order to put tighter constraints from the analyses. In other words, we search for deviations from CDDR validity by using the relations presented in the previous section (see Eq.\ref{rel}). The samples used here are:

\begin{itemize}
\item We consider 29 well-described galaxy clusters by a spherical non-isothermal double $\beta$ model from a original sample of 38 from Ref.\cite{bonamente}, see Fig.(1a). This model take into account a possible presence of cooling flow in galaxy cluster cores. We cut-off the galaxy clusters that presented questionable reduced $\chi^2$ ($2.43 \leq \chi^2{d.o.f.} \leq 41.62$) when described by the hydrostatic equilibrium model. It is important to stress that the frequency used to obtain the SZE signal in galaxy clusters sample considered was 30 GHz, in this band the effect on the SZE from a variation of $T_{CMB}$ is completely negligible. The best frequency is 150 GHz for negative signals and around 260 GHz for positive signals. Therefore, we do not consider a modified CMB temperature evolution law in the galaxy cluster data.

\item The full SNe Ia sample is formed by 580 SNe Ia data compiled in Ref.\cite{suzuki}, the so-called Union2.1 compilation, see Fig.(1a).  In order to perform our test we need SNe Ia and galaxy clusters in the identical redshifts. In this way, we consider the 29 angular diameter distance of galaxy clusters from the sample of \cite{bonamente} and,  for each i-galaxy cluster, we obtain one  distance modulus, $\bar{\mu}$, and its error, $\sigma^2_{\bar{\mu}}$, from all i-SNe Ia  with $|z_{cluster_i} - z_{SNe_i}| \leq 0.006$. Naturally, this criterion allows us to have some SNe Ia for each galaxy cluster and so we can perform  a  weighted average with them in order to minimize the scatter observed on the Hubble diagram. Then, we calculate the following weighted average \cite{meng} from SNe Ia data:
\begin{equation}
\begin{array}{l}
\bar{\mu}=\frac{\sum\left(\mu_{i}/\sigma^2_{\mu_{i}}\right)}{\sum1/\sigma^2_{\mu_{i}}} ,\hspace{0.5cm}
\sigma^2_{\bar{\mu}}=\frac{1}{\sum1/\sigma^2_{\mu_{i}}}.
\end{array}\label{eq:dlsigdl}
\end{equation}
\item The $T_{CMB}(z)$ sample is composed by 38 points, see Fig.(1b). The data in low redshift are from SZE observations \cite{luzzi} and from observations of spectral lines we have the data at high redshift \cite{hurier}. In total, this represents 38 observations of the CMB temperature at redshift between 0 and 3. We also use the estimation of the current CMB temperature $T_0 = 2.725 \pm 0.002$ K \cite{mather} from the CMB spectrum as estimated from the COBE collaboration (see Fig.(1b)).
\end{itemize}
\begin{figure*}[ht]
\centering
\includegraphics[width=0.47\textwidth]{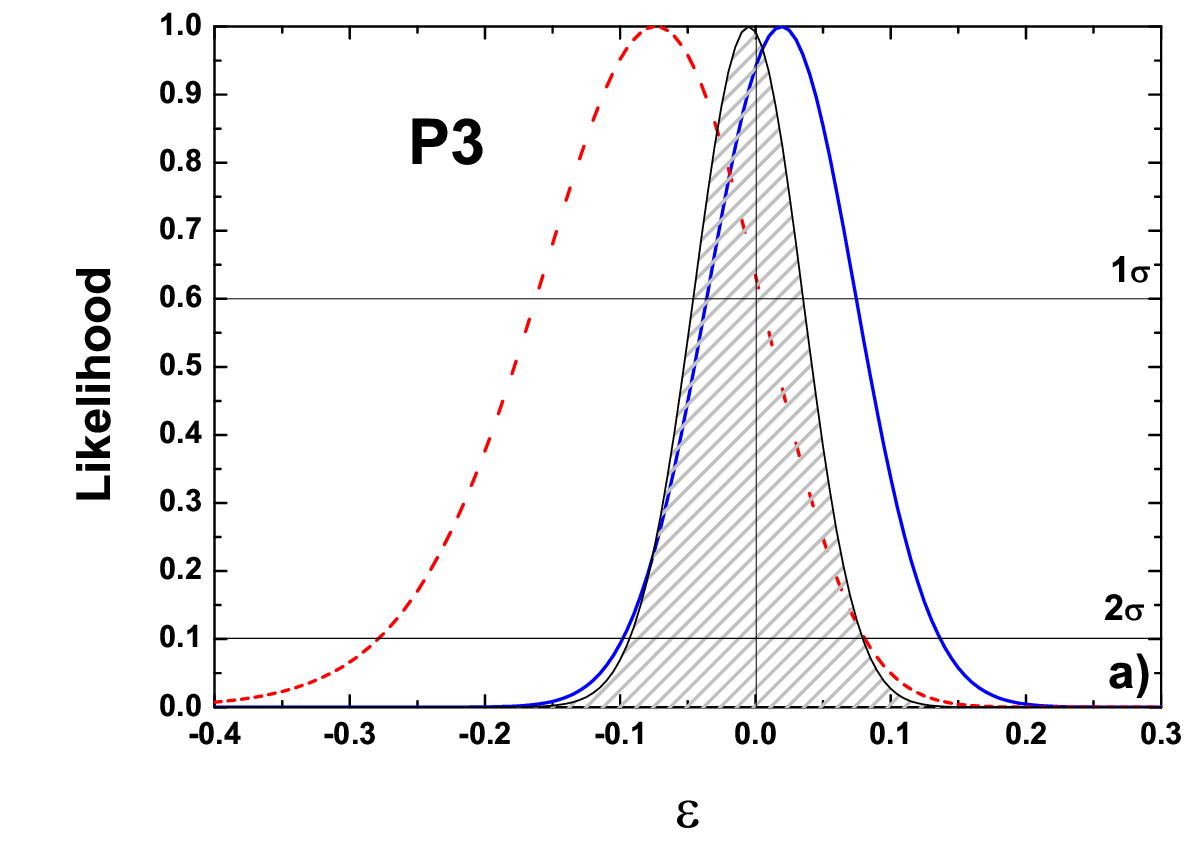}
\hspace{0.3cm}
\includegraphics[width=0.47\textwidth]{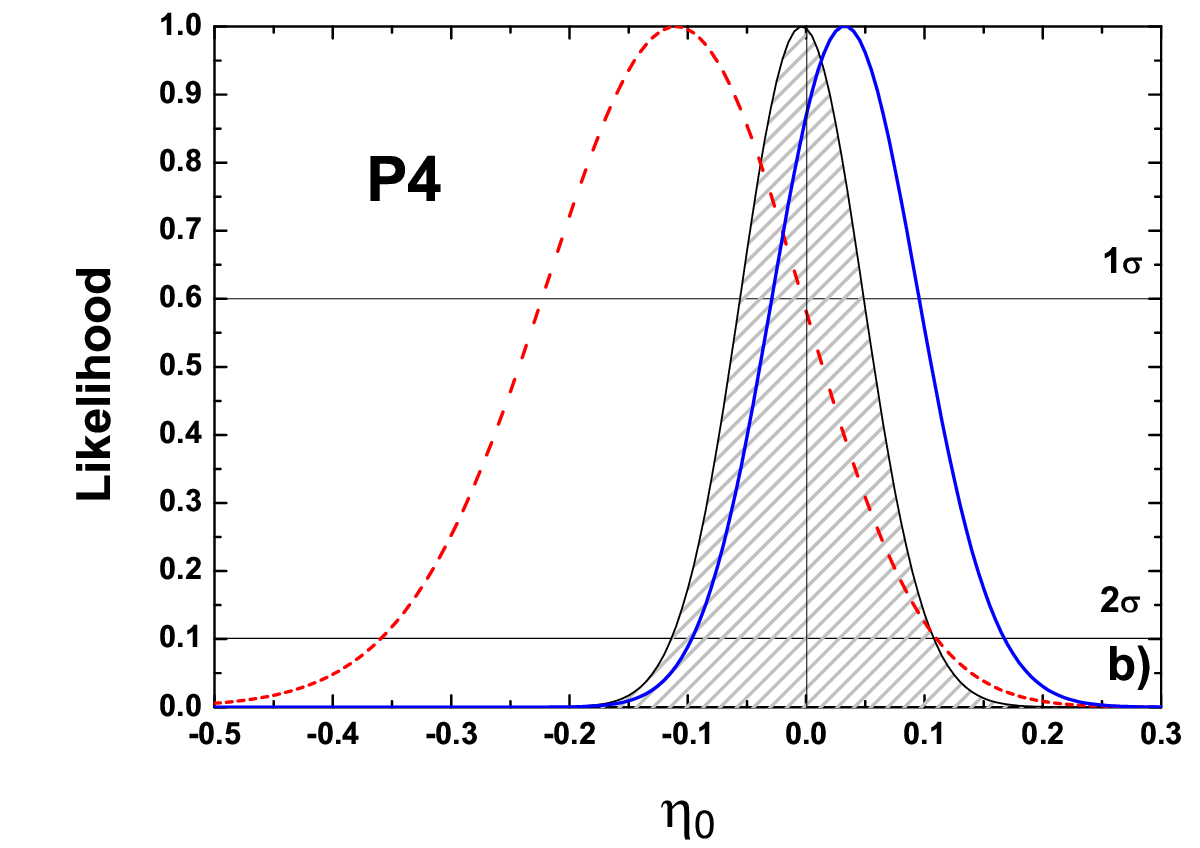}
\caption{In both figures, the solid blue and dashed red lines correspond to analyses by using SNe Ia + GC and $T_{CMB}(z)$, respectively. The dashed area corresponds to the joint analysis (SNe Ia + GC + $T_{CMB}(z)$). In Fig.(a) we plot the results by using the parametrization P3 and in Fig.(b) by using P4.}
\end{figure*}
\section{Analyses}
As already discussed in  \cite{holandaprd}, for the class of theories discussed by Hees et al., the SZE + X-ray measurements of galaxy clusters do not give the true ADD, but $D_A^{data}=\eta^4(z)D_A$ (where was used $\phi=\eta^2$ in Eq.(\ref{eqnew}).  Moreover, as argued in \cite{uza}, if one wants to test the CDDR  by using $D_L(1+z)^{-2}D_A^{-1}=\eta$ and galaxy clusters via SZE/X-ray technique, the ADD $D_A(z)$ must be replaced by $D_A(z)=\eta^{-4}D_A^{data}$ ($\eta^{-2}$ in their case, since variations of $\alpha$ were not considered). In this way, we have access to 
\begin{equation}
\frac{D_L}{(1+z)^2D_A^{data}(z)}=\eta^{-3}(z).
\end{equation}
By using the equation above, we define the distance modulus of a galaxy cluster data as
\begin{equation}
 \mu_{cluster}(\eta ,z)=5\lg[\eta^{-3}(z)D^{data}_A(z)(1+z)^2]+25. 
\end{equation}
We evaluate our statistical analysis by defining the likelihood distribution function, ${\cal{L}} \propto e^{-\chi^{2}/2}$, where 
\begin{eqnarray}
\chi^2 &=& \sum_{i=1}^{29}\frac{(\bar{\mu}(z_i)-\mu_{cluster}(\eta,z_i))^2}{\sigma_{obs}^2}\nonumber \\ &&+ \sum_{i = 1}^{38}\frac{{\left[ T(z_i) - T_{i,obs} \right] }^{2}}{\sigma^{2}_{T_i, obs}} ,
\label{chi}
\end{eqnarray} 
with $\sigma_{obs}^2= \sigma^2_{\bar{\mu}} + \sigma^2_{\mu cluster}$ and  $T(z)$  given by Eq.(8). The sources of statistical uncertainty in the error bars of $D_{A}^{\: data}$(z)  are:  SZE point sources $\pm 8$\%, X-ray background $\pm 2$\%, Galactic N$_{H}$ $\leq \pm 1\%$, $\pm 8$\% kinetic SZ and for CMB anisotropy $\leq \pm 2\%$. We have added in quadrature the following systematic errors: SZ calibration $\pm 8$\%, X-ray flux calibration $\pm 5$\%, radio halos $+3$\%, and X-ray temperature calibration $\pm 7.5$\%. Following  \cite{suzuki} we added  a 0.15 systematic error to SNe Ia data.

In order to explore the dependence of our results with $\eta(z)$ function, we consider 4 parametrizations, namely:
\begin{itemize}
\item P1: $\eta(z)=1+\eta_0 z$

\item P2: $\eta(z)=1+\eta_0 z/(1+z)$

\item P3: $\eta(z)=(1+z)^{\epsilon}$

\item P4: $\eta(z)=1+ \eta_0 \ln(1+z)$
\end{itemize}
where $\eta_0$ and $\epsilon$ are the parameters to be constrained. The limits $\eta_0=\epsilon=0$ corresponds to the standard GR results.

Our results are plotted in Figs. (2) and (3) for each parametrization and samples described in Sec.III. {Note that in each case the  solid (blue) and dashed (red) lines correspond to analyses by using separately  CMB temperature and galaxy clusters + SNe Ia data in Eq.(17), respectively. The dashed area are the results from the joint analysis, i.e., the complete Eq.(17) with CMB temperature + galaxy clusters + SNe Ia. In Table I we put our 1$\sigma$ results from the joint analyses for each parametrization and several $\eta_0$ values present in literature, obtained by using ADD from galaxy clusters + SNe Ia that did not take into account the effect of a possible $\alpha$ variation on the SZE/X-ray technique. Notoriously our results present tighter limits on $\eta_0$ value than previous analyses.} We also present the results obtained by using the galaxy clusters sample from Ref.\cite{fil}, where the X-ray surface brightness was described by a elliptical isothermal $\beta$-model  in order to compare results. In this case, the galaxy clusters are distributed over the redshift interval $0.023 \leq z \leq 0.784$. It is very important to consider another assumptions on the galaxy clusters morphology since the ADD depends on the hypotheses considered.  As one may see, our results are in full agreement each other regardless the galaxy clusters sample and $\eta(z)$ function used. Moreover, no indication of deviations from standard results is obtained. 
\begin{table*}[ht]
\caption{A summary of the current constraints on the parameters $\eta_0$ for P1, P2 and P4 and $\epsilon$ for P3,  from angular diameter distance from galaxy clusters and different SNe Ia samples. The symbol * Corresponds to angular diameter distance (ADD) from Ref.\cite{fil} and ** angular diameter distance from Ref.\cite{bonamente}. The symbol $\ddag$ corresponds to analyses which do not consider variations of $\alpha$ on the SZE/X-ray technique.}
\label{tables1}
\par
\begin{center}
\begin{tabular}{|c||c|c|c|c|c|}
\hline\hline Reference & Data Sample &$\eta_0$ (P1)& $\eta_0$ (P2)& $\epsilon$ (P3)& $\eta_0$ (P4)
\\ \hline\hline 
\cite{holanda2011} & $ADD^{\ddag *}$  + SNe Ia  &$-0.28 \pm 0.24$ & $-0.43 \pm 0.21$ & - &- \\ 
\cite{holanda2011} & $ADD^{\ddag **}$  + SNe Ia  &$-0.42 \pm 0.14$ & $-0.66 \pm 0.16$ & - &- \\
\cite{li}& $ ADD^{\ddag *}$  + SNe Ia & $-0.07 \pm 0.19$ & $-0.11 \pm 0.26$ & - &-\\
\cite{li}&  $ADD^{\ddag **}$  + SNe Ia & $-0.22 \pm 0.11$ & $-0.33 \pm 0.16$ & - &-\\
\cite{liang}& $ADD^{\ddag **}$  + SNe Ia& $-0.23 \pm 0.12$ & $-0.35 \pm 0.18$ & - &-\\ 
\cite{meng} & $ADD^{\ddag *}$  + SNe Ia& $-0.047 \pm 0.178$ & $-0.083 \pm 0.246$ & - &-\\
\cite{meng}  & $ADD^{\ddag **}$  + SNe Ia& $-0.201 \pm 0.094$ & $-0.297 \pm 0.142$ & - &-\\
 \cite{yang} & $ADD^{\ddag *}$   + SNe Ia      & $0.16^{+0.56}_{-0.39}$    & - & - &-\\
\cite{yang} & $ADD^{\ddag **}$   + SNe Ia      & $0.02 \pm 0.20$    & - & - &-\\ 
\cite{holandaprd} & $ADD^{*}$   + SNe Ia      & $0.069 \pm 0.106$    & 0 $\pm. 0.135$ & - &-\\
\bf{This paper} & $ADD^{**}$   + SNe Ia  + $T_{CMB}$ & $-0.005 \pm 0.025$& $-0.048 \pm 0.053$ & $-0.005\pm 0.04 $&$-0.005 \pm 0.045$ \\
\bf{This paper} & $ADD^{*}$   + SNe Ia  + $T_{CMB}$  & $-0.005 \pm 0.032$& $-0.007 \pm 0.036$ &$ 0.015 \pm 0.045$ & $0.015 \pm 0.047$\\
\hline\hline
\end{tabular}
\end{center}
\end{table*}
\section{Conclusion}
In Ref.\cite{hees} it was developed a powerful mechanism to test  signatures of a wide class of theories of gravity that explicitly breaks the Einstein equivalence principle. Briefly, they introduced an additional term into the action (see Eq.(1)) coupling the usual matter fields to a scalar additional field, which is motivated by scalar-tensor theories of gravity, for instance. Actually, all the electromagnetic sector of the theory  is affected,  leading to deviations of the CDDR validity, $D_L(1+z)^2/D_A=\eta$, variations of fundamental constants, $\Delta \alpha/\alpha$ (where $\alpha$ is the fine structure constant), and  of the evolution law of the Cosmic Microwave Background radiation. The distortions of the standard results are related by Eq.(11). 

In this paper, we have used ADD of galaxy clusters obtained via their Sunyaev-Zeldovich effect (SZE) + X-ray surface brightness observations, luminosity distances of SNe Ia and CMBR temperature to search signatures of the class of theories considered. By properly considering these deviations in the data, mainly on the SZE/X-ray technique which depends explicitly on the $\eta$ and $\alpha$, we put constraints on four parametrization of $\eta(z)$ via a jointly analysis of data (see last two lines of Table I). {We have obtained  tighter constraints on possible deviations from GR than previous works, and  all case were found to be in full agreement with standard GR framework, $\eta=1$. However, the results presented here do not  rule out  the models under question with high confidence level yet. When larger samples with smaller statistical and systematic uncertainties of  X-ray and SZE observations as well as $T_{CMB}(z)$ measurements and SNe Ia become available, the method proposed here will be able to search deviations from the standard framework with more accuracy. }
\begin{acknowledgements}
RFLH acknowledges financial support from  CNPq and UEPB (No. 478524/2013-7, 303734/2014-0). SHP is grateful to CNPq - Conselho Nacional de Desenvolvimento Cient\'ifico e Tecnol\'ogico, Brazilian research agency, for financial support, grants number 304297/2015-1. 
\end{acknowledgements}

\end{document}